# Diffraction Patterns of Apertures Shaped as National Borders

Albert F. Rigosi[1*]

[1]*Physical Measurement Laboratory, National Institute of Standards and Technology (NIST), Gaithersburg, MD 20899, United States*

[2]*Department of Physics, Columbia University, New York, NY 10027, United States*

ABSTRACT: How aesthetically pleasing is your country's diffraction pattern? This work summarizes the calculated and experimental Fraunhofer diffraction patterns obtained from using apertures lithographically formed into shapes of national borders. Calculations are made based on the fast Fourier transform of the aperture images. The entropy of each of the 113 nations' diffraction patterns was also computed based on its two-dimensional gradient. Results suggest that most nations' diffraction patterns fall under one of two prominent trends forming as a function of geographical area, with one trend being less entropic than the other.

[*] Email: albert.rigosi@nist.gov

Have we ever considered how elegant a diffraction pattern may appear based on an aperture in the shape of one's nation? One of the many interesting facets of basic electromagnetic theory is the propagation of light through a well-defined aperture. Propagation of electrons may also be considered in mediums such as graphene, where ballistic behavior enables light-like responses to changes in material environment (such as *p-n* junctions) [1, 2]. At short distances after the aperture, specifically those within the same order of magnitude as the aperture size, the interference of light falls under the regime of Fresnel, or near-field, diffraction. On the other hand, when observing interference effects at distances much greater than the aperture size, a well-defined pattern emerges, falling into the Fraunhofer, or far-field, diffraction regime. To consider the diffraction of light in the far-field regime, one can simplify the Fresnel–Kirchhoff integral theorem listed as:

$$U_P(x,y,z) = \frac{1}{4\pi} \iint \left[ U \frac{\partial}{\partial n}\left(\frac{e^{i\frac{2\pi}{\lambda}s}}{s}\right) - \left(\frac{e^{i\frac{2\pi}{\lambda}s}}{s}\right) \frac{\partial U}{\partial n} \right] dS$$

(1)

In Equation 1 above, $U_p$ represents the complex amplitude of the optical disturbance, or the light wave's electric field, at a point *P* after the aperture. By assuming that we are both in the far-field regime and that the incoming and outgoing light waves are plane waves parallel to the aperture, then one obtains:

$$U_P(x,y,z) \propto \iint A(x',y') e^{i\frac{2\pi}{\lambda}(lx'+my')} dx'\, dy'$$

(2)

Equation 2 is the simplification of the Fresnel–Kirchhoff integral theorem when considering far-field effects only, and the function $A(x', y')$ is essentially the aperture area. There have been many efforts to evaluate the diffraction of various unconventional apertures, and some of those were experimental in nature [5-8], whereas others were focused on the numerical or analytical evaluation of more complex apertures [9-12]. Analytical solutions for apertures resembling national borders would not be possible given the intricate parameterization required to describe the exact shape of the aperture. Instead, we focus on the experimental and numerical analysis of these diffraction patterns.

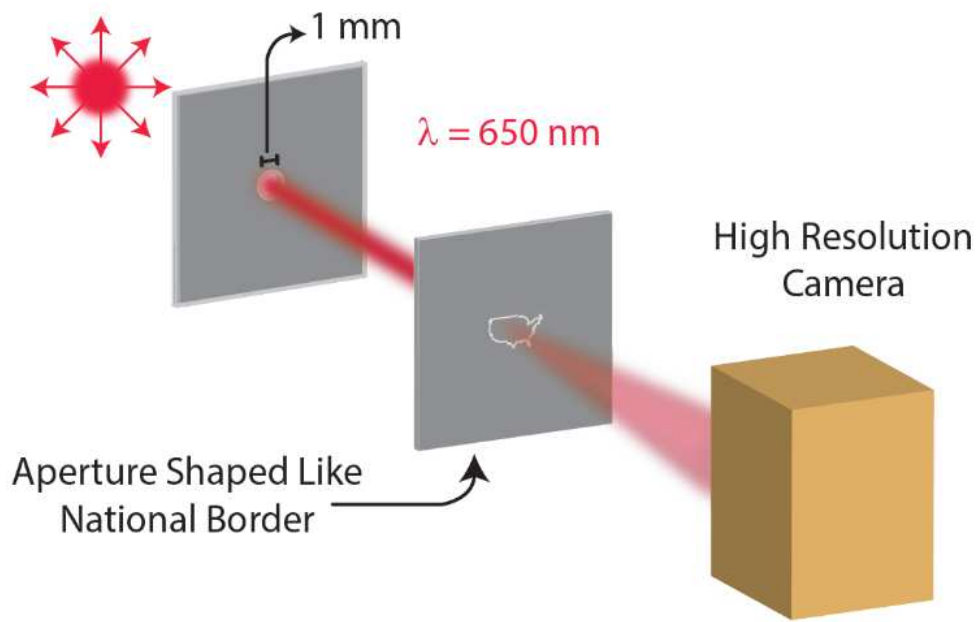

FIG. 1. A simplified illustration of the experimental setup is depicted here. A 650 nm laser source has a restricted output diameter of 1 mm and is routed through an aperture having the shape of one of 113 national borders. The Fraunhofer diffraction pattern is directed to a high resolution camera and recorded at various exposure times depending on the aperture.

To set up the experiment, a 650 nm laser diode source (from US Lasers, Inc) [3] with a 5 nm tolerance and a 5 mW power was used and filtered through a 1 mm aperture to remove any irregular beam profiles inherent to the source. The laser was routed approximately 10 cm before being directed to the main aperture of interest. The initial 1 mm aperture had caused a small amount of diffraction of the laser diode source, but the fringes were separated by several centimeters when the source reached the aperture of interest, thus removing possible interactions between the fringes and measured apertures. The resulting diffraction pattern was recorded with a Nikon DS-Ri2 16.25 megapixel microscope camera, attached to the MM-400 microscope [3].

All apertures were fabricated on quartz substrates with a thickness of 170 μm by first depositing 350 nm of chromium. Photolithography was then used to etch the national borders, which scaled proportionately with the geographical area of each used nation. The apertures were written into a vector file format with national borders based on Gall stereographic projection.

The nations were selected in order of total area, as provided by the United Nations Statistics Division, starting with the largest nation and including all nations with an area larger than $8 \times 10^{10}$ m$^2$. Only mainland nations included in the International Organization for Standardization (ISO) document ISO 3166-1 were used (thus excluding Danish Greenland, the Republic of Somaliland, and the disputed territory of Western Sahara).

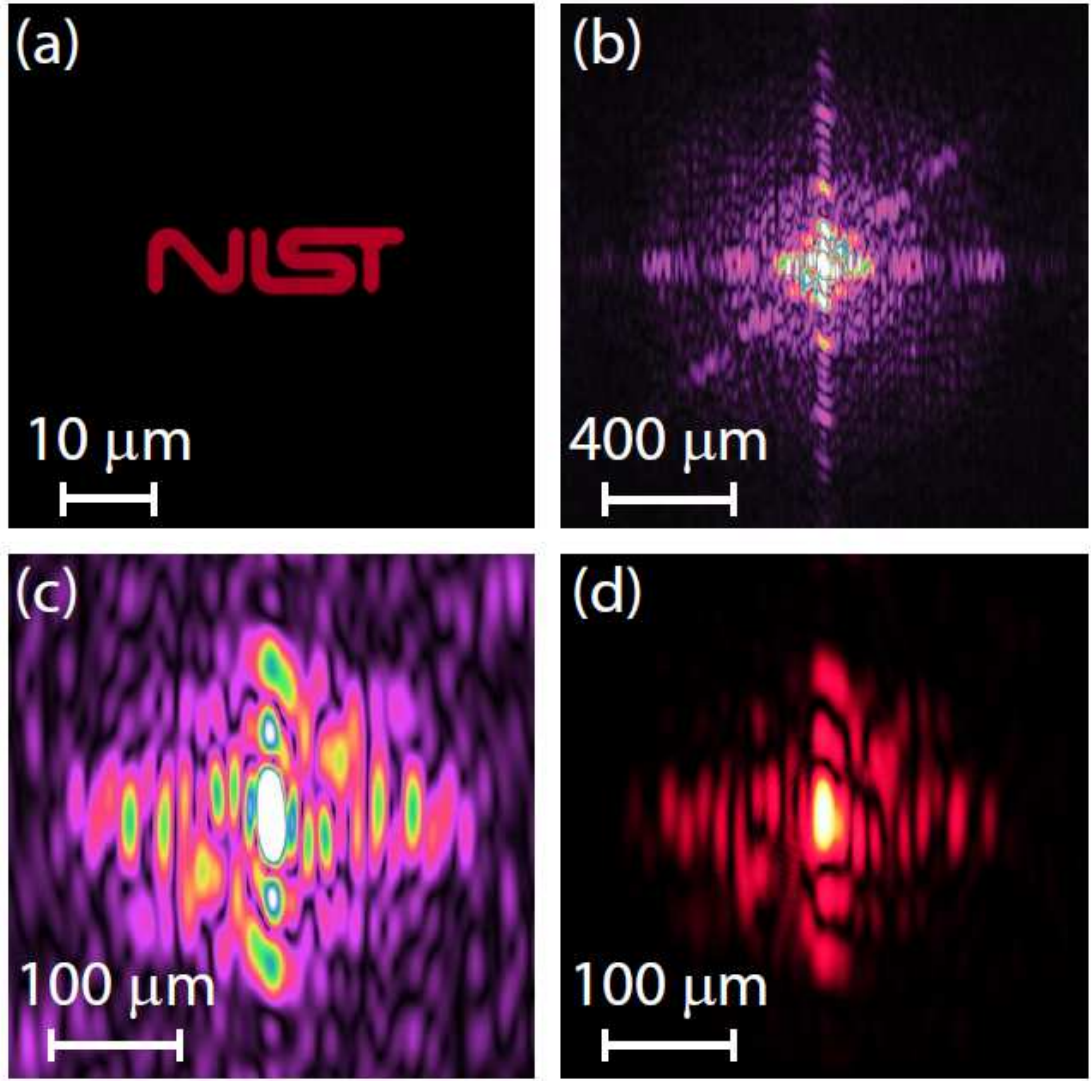

FIG. 2. An example analysis is shown here comparing experimental data to the calculated the fast Fourier transform (FFT) of an example aperture. (a) The aperture was etched using photolithography methods and is shaped like the logo for the National Institute of Standards and Technology (NIST). The image was taken while the laser source was on. (b) The FFT of (a) shows the expected diffraction pattern as it would appear on a focal plane 1.35 mm away from the aperture. (c) When magnifying the center region by a factor of 4, it was easier to compare with the experimental data in (d), which was also taken at 1.35 mm away from the aperture.

It is important to note that the form of Equation 2 is that of a Fourier transform of a spatial function representing the coordinates of the aperture. This form also enables one to make straightforward calculations of the predicted diffraction patterns for a variety of apertures. *Gwyddion* open-source software was used for calculating the fast Fourier transforms (FFTs) of real-space images of the apertures after they were developed [4]. An example analysis is summarized in Figure 2, with the aperture of choice being the logo for the National Institute of Standards and Technology in the United States (USA).

The image shown in Fig. 2 (a) was taken while the aperture was being illuminated by the laser diode source. The FFT was calculated in (b) with the scale corresponding to a distance of 1.35 mm away from the aperture. The experimental data for the Fraunhofer diffraction pattern is displayed in (c). For a pattern like the one in (b), the power required for observing it would be on the order of 100s of mW. This order-of-magnitude drop-off in observed power is expected of most classical diffraction patterns [1, 13]. For safety and to reduce possible damage to the microscope camera, only 5 mW was used to illuminate the magnified region. The effective intensity, or power density, was identical for all apertures.

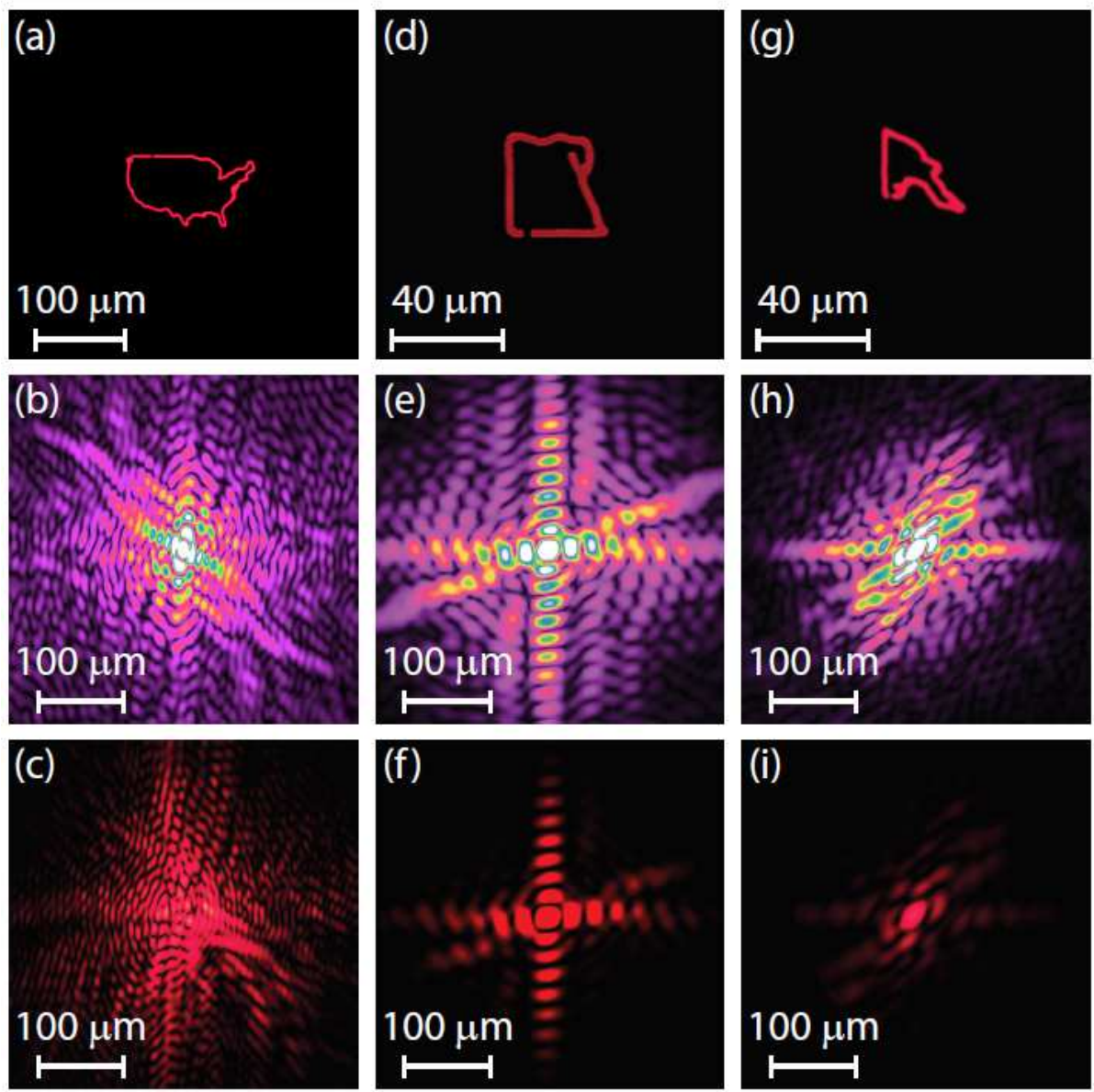

FIG. 3. Three example larger nations are presented here. (a) The aperture for the continental USA is depicted. (b) is the FFT calculation of the aperture above, and the corresponding experimental data is shown below in (c). (d) The aperture for Egypt is depicted, along with its FFT and experimental data in (e) and (f), respectively. (g) The aperture for Papua New Guinea is shown with its (h) calculated FFT and (i) experimental diffraction data.

With this example analysis in mind, the data for each of the 113 nations' apertures was systematically collected and analyzed, with a subset of the larger nations shown in Figure 3. Fig. 3 (a) shows the aperture representing the continental United States of America. Due to its large size relative to other nations, the Fraunhofer diffraction is more disordered, as can be seen in its calculated FFT in (b), and compared to with experimental data in (c). In Fig. 3 (d), the aperture representing the Arab Republic of Egypt was analyzed. The lateral scale bars were adjusted to reflect the proportionate sizes of the apertures relative to each other. The calculated FFT of the aperture and experimental data for Egypt's diffraction pattern are shown in (e) and (f), respectively. One could differentiate the pattern by noticing the two strong axes accompanied by a third axis at a low-sloped diagonal. The third aperture represents the Independent State of Papua New Guinea in Fig. 3 (g) and its calculated FFT and corresponding experimental diffraction data are displayed in (h) and (i), respectively. It can be distinguishing by its resemblance to a spiral galaxy.

  A different subgroup of nations with smaller geographical areas were selected for display in Figure 4 to notice the trends and changes in the Fraunhofer diffraction patterns. Fig. 4 (a) displays the aperture representing the State of Japan, whereas (d) displays the aperture for the Italian Republic. For both cases, as seen in the FFT and data in (b), (c), (e), and (f), the diffraction pattern takes on a more rippled appearance. The final nation to be analyzed here in detail was the Federal Democratic Republic of Nepal, whose aperture is shown in Fig. 4 (g), along with its corresponding FFT calculation and experimental data in (h) and (i), respectively. Ranking 93 out of 113 in geographical size, Nepal's diffraction pattern appeared to have far less disorder than the other two nations.

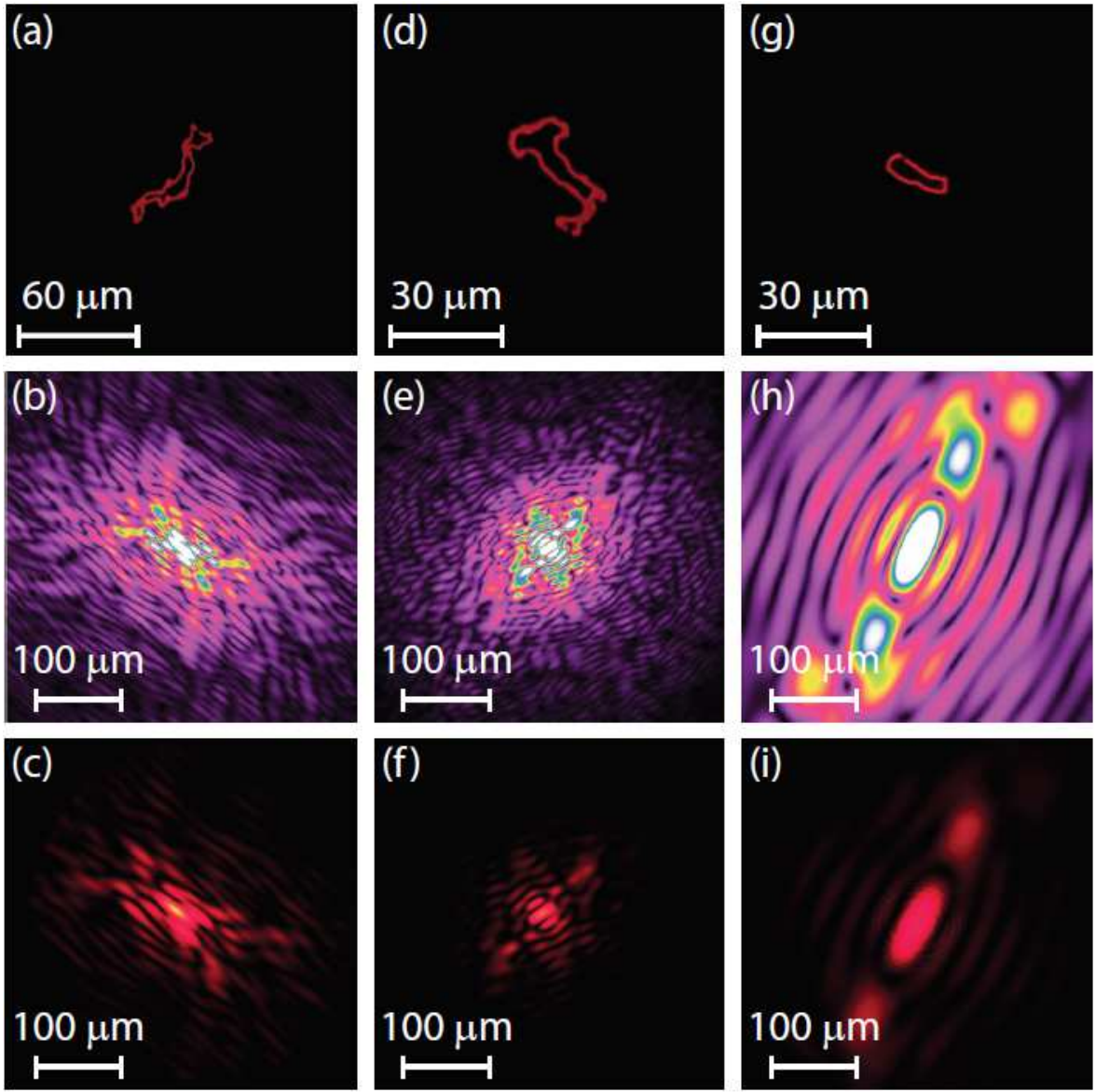

FIG. 4. Three example smaller nations are presented here. (a) The aperture for Japan is depicted. (b) is the FFT calculation for Japan's aperture above, and the corresponding experimental data is shown below in (c). (d) The aperture for Italy is shown along with its FFT and experimental data in (e) and (f), respectively. (g) The aperture for Nepal is shown and its calculated FFT and experimental diffraction data are presented in (h) and (i), respectively.

Following the collection of 113 different Fraunhofer diffraction patterns [14], a quantification of the level of disorder present in the predicted FFT images was sought. Entropy, the statistical measure of randomness, was used to characterize the image disorder, which can be done with standard formulas [15, 16]. One interpretation of the entropy is the lowest possible average number of bits per pixel. The FFT images, which were only scaled in color for presentation, were used in grayscale for these calculations, where every pixel can use up to eight bits (or 256 shaded tones) to encode information about intensity. The entropy is defined as:

$$S = \sum_i p_i \log p_i$$

(3)

In equation 3, $p$ represents the number of counts per pixel $i$ in a normalized histogram of intensities of the image. One major problem in determining the entropy of the image is that a simple calculation, such as the one provided with the *entropy* function in MATLAB [3], will not consider the long-range order or disorder in the total entropy [17]. For instance, Larkin shows that if two images are composed of the same number of pixels for each shade, the entropies will be similar [17]. An easy way to notice this problem is by taking the example of the following images: one white noise image, and one with a single-direction gradient (black on one side and gradually going to white on the other). Both have the same entropy based on the simple *entropy* function or equation 3. It would be more intuitive to associate a higher entropy for white noise as opposed to a gradient.

This problem of gaining better insight into the diffraction pattern disorder can be resolved by considering the entropy of the pattern's two-dimensional gradient [17]. If one considers the two-dimensional gradient of the diffraction pattern image, a more intuitive entropy for the original image can be obtained. With the examples given in the previous paragraph, the entropy of the white noise image remains large since taking its two-dimensional gradient does not mask any hidden order. However, a grayscale gradient, when differentiated, has a constant image, and therefore has a very small entropy.

The image entropy was analyzed with the method of using two-dimensional gradients as described above. The results shown are shown in Figure 5. Example two-dimensional gradients are displayed in Fig. 5 (a-d) for the FFTs of the USA, Kingdom of Spain, Italy, and Nepal apertures. These gradients are then used for entropy calculations, revealing a more accurate behavior for the group of FFTs. In Fig. 5 (e), the calculated diffraction pattern entropy is shown as a function of the ranking order of nations' areas. A fitted line in teal is also provided for the data, suggesting that the diffraction pattern entropy increases with smaller nations.

However, one issue with evaluating entropy based on the $1 \times$ magnification of the FFT, like the one shown in Fig. 2 (b), is that some information on the subtler intricacies of the diffraction pattern is lost when the magnification is too low. This loss is a result of image resolution, which can be remedied by focusing on a higher magnification of the FFT. Lastly, the increasing trend in the entropy makes sense only for this magnification because smaller nations will have apertures that diffract light into generally wider patterns, thus appearing to have larger entropy in the limit of lower image resolution.

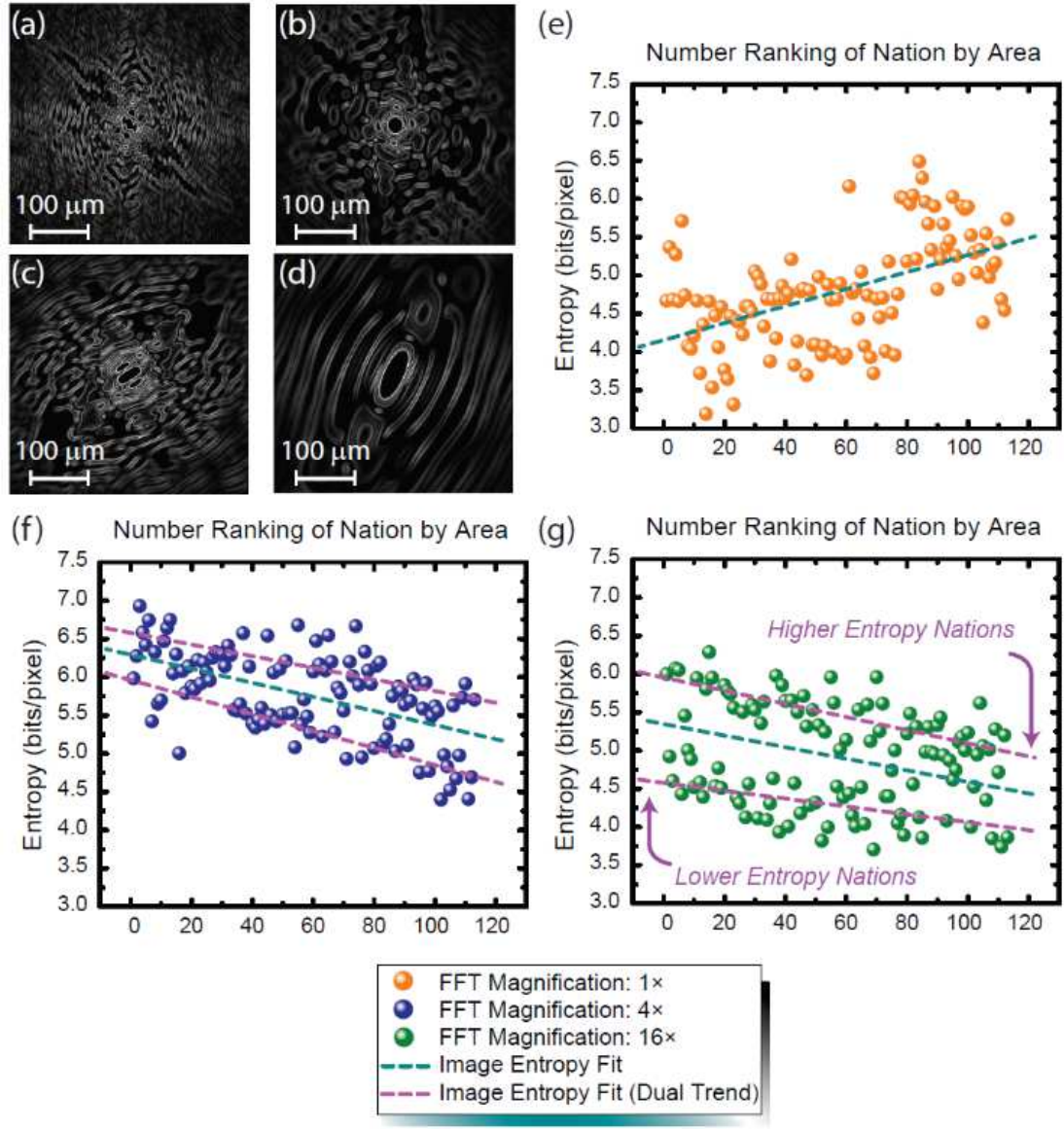

FIG. 5. An analysis of the entropy of the Fraunhofer diffraction images is summarized here. Two-dimensional gradient calculations are provided for the diffraction patterns from the (a) USA, (b) Spain, (c) Italy, and (d) Nepal apertures. The gradient images provide a better insight into the entropy of the original diffraction images. (e) The final entropies of the diffraction pattern calculations (FFTs) are shown with no magnification, as in Fig. 2 (b). A dotted teal line shows the linear fit to the data. (f) The FFTs are analyzed at a higher magnification factor (4 ×) to reveal more detail and resolution in the image. The data is fit to a teal line, with the addition of a second set of linear fits in purple which draws attention to the formation of two

subgroupings of nations with noticeably different entropies. (g) The FFTs are analyzed at a higher magnification (16 ×) and are plotted in a similar to fashion to (f). In this panel, the two trends are labelled for clarity.

To extend the analysis to magnified FFT regions, the scaling factors 4 × and 16 × were used to refine the FFTs, thus reducing the amount of lost data towards the center of the diffraction pattern. The 4 × and 16 × magnified FFTs were evaluated for the entropic trends in Fig. 5 (f) and (g), respectively. The teal dotted lines are linear fits to all of the data.

There are two points to with these analyses. The first point is that the entropy trends have reversed such that smaller nations' apertures generate less entropic diffraction patterns. This behavior does make sense, since in the limit of magnification to the center of the FFT, most of the diffraction patterns for smaller nations lie outside of the bounds of the calculated image, and with that, so does the entropy.

The second point is that both magnifications appear to accommodate two separate linear trends, which are marked as purple dashed lines in Fig. 5 (f) and (g). As the magnification on the FFT region increases, the entropy has a negative slope and splits into two distinct entropic trends, one represented by higher entropy nations, and the other by lower entropy nations. The trends temporarily diverge in the 4 × picture and begin to converge in the 16 × picture. This resulting analysis thus implies that, should a nation wish to reduce their Fraunhofer diffraction image entropy, the border should change into a shape whose diffraction pattern has more gradual changes in intensity, corresponding to a less chaotic two-dimensional gradient.

In summary, the calculated and experimental Fraunhofer diffraction patterns were obtained by using apertures matching the shapes of national borders. The entropy of those diffraction pattern images for 113 nations has also been computed, with the results suggesting that two trends, ones of generally higher and lower entropy, appear and that most nations' diffraction patterns more clearly fall into one of the two entropic trends. One interesting application or possible extension to this work would be to attempt such a diffraction pattern using electrons All nations' apertures, FFT calculations, and experimental data are provided in [14]. Nations marked with the letters from A through F are presented earlier in this work [18]. All calculations and data are presented from a perspective similar to the main text, that is, 1.35 mm away from the aperture.

## ACKNOWLEDGMENTS AND NOTES

AFR thanks the National Research Council, Ford Foundation, and National Science Foundation for the opportunity. Funding is acknowledged from the National Science Foundation (NSF DGE-1144155).